\documentclass[runningheads,a4paper]{llncs}

\usepackage{mathptmx}       
\usepackage{helvet}         
\usepackage{courier}        
\usepackage{multicol}        
\usepackage[bottom]{footmisc}

\makeindex             

\usepackage{latexsym}
\usepackage{proof}
\usepackage{alltt}

\newcommand{\bs}{\backslash}

\begin{document}

\title{A Conjecture for ATP Research}
\author{Wolfgang Bibel}
\institute{Darmstadt University of Technology,\\
  \email{bibel@gmx.net}, \today}

%

\maketitle

\abstract{This note generalizes factorization for formulas with multiplicities and conjectures that the connection method along with this feature is computationally as powerful as resolution, also seen from a complexity point of view.}

%


\section{Factorization Revisited}

This short note revisits a particular aspect of factorization within the Connection Method (CM) and derives on its basis a conjecture stated in the subsequent section along with possibly far reaching consequences for the field of Automated Theorem Proving (ATP), in case the conjecture can formally be established as true.

We assume familiarity with the terminology underlying the CM (see~\cite{bi:B8} and/or any of the respective publications on CM published thereafter). This includes the term {\em factorization\,}~\cite[Sect.IV.6]{bi:B8}. In short, it denotes exploiting during proof search the situation of an identical literal occurring at different positions in a formula.

Such a situation may happen also in formulas with multiplicities~\cite[Sect.III.6, 6.1.D]{bi:B8} which is the focus of the present note. We illustrate this case with the following example.
\vspace*{-3mm}

\begin{figure}
\begin{center}
\vspace*{2mm}
\begin{minipage}{8cm}
		$N0\wedge\forall x(Nx\rightarrow Nfx)\rightarrow Nff0\wedge Nf0$ ~ $\sigma_1 = \{x_1\bs 0, x_2\bs f0, x_3\bs 0\}$

{\setlength{\unitlength}{0.25mm}
\begin{picture}(0.00,0.00)(0.00,0.00)
\qbezier(10.00,28.00)(40.00,45.00)(50.00,28.00)
\put(50,30){\scriptsize\bf 1}
\qbezier(50.00,11.00)(68.00,-10.00)(85.00,11.00)
\put(54,4){\scriptsize\bf 2}
\put(77,4){\scriptsize\bf 1}
\qbezier(87.00,28.00)(98.00,45.00)(128.00,28.00)
\put(83,30){\scriptsize\bf 2}
\qbezier(10.00,11.00)(40.00,-10.00)(50.00,11.00)
\put(44,4){\scriptsize\bf 3}
\qbezier(87.00,11.00)(128.00,-10.00)(165.00,11.00)
\put(86,4){\scriptsize\bf 3}
\end{picture}
}
\end{minipage}
\vspace*{-5mm}
\end{center}
\caption{A connection proof for formula $F_1^{\mu_1}$.\label{fig:ConnProofF1}}
\end{figure}

\vspace*{-4mm}
The figure shows a connection proof for a simple formula $F_1^{\mu_1}$ with multiplicity $\mu_1$ along with the unifying substitution. Connections thereby are illustrated with arcs as usual; formally these are pairs of occurrences of literals along with the corresponding multiplicity. The multiplicity of the ground literals is $1$ by default. Recall that connection proofs for any such formula consist of an associated multiplicity and a spanning set of connections with a unifying substitution~\cite[6.4.C]{bi:B8} as illustrated in the figure. The shown proof might be regarded as the standard one for this example.

Assume that the occurrences of the five literals in the formula are denoted by $0,1,\ldots,4$. Then the leftmost upper connection formally is $c_1^{11}=(0^1,1^1)$ whereby the multiplicity is attached to the occurrence as an upper index. Similarly for the other four connections $c_2^{13},c_3^{21},c_4^{21},c_5^{31}$ listed top down and from left to right.

Now we may note that $Nx_1\sigma_1=Nx_3\sigma_1$ and thus $c_1^{11}=c_2^{13}$, ie.\ these two literals may be factorized rendering one of the two connections as superfluous; one might even speak of factorizing these two connections. Similarly, $Nfx_1\sigma_1=Nfx_3\sigma_1$ and thus these two literals may be factorized as well. In consequence, the proof now for $F_1^{\mu_1'}$, with a different, in fact slightly simplified multiplicity $\mu_1'$, is in fact established with just the four connections $c_1^{11},c_3^{21},c_4^{21},c_5^{11}$ and with the substitution $\sigma_1' = \{x_1\bs 0, x_2\bs f0\}$. Hence factorization may yield smaller connection proofs as illustrated with this example.

This simple example thus illustrates that this kind of {\em factorization of formulas with multiplicities\,}, as we refer to it, may simplify connection proofs as illustrated with this example. In consequence, it may also speed up proof search. Formally, we express factorization in this example as $\varphi_1 F_1^{\mu_1}$, whereby $\varphi_1$ denotes the particular factorization just described and $\mu_1$ the multiplicity involved with $F_1$. The concept is an obvious one, also in its full generality, and simply consists in a variation of the familiar concept of factorization in ATP.

It is amazing that the present author is not aware of this kind of factorization having been taken notice in any pertinent publication during more than the past half a century. This is further evidence for the statement made in~\cite[Sect.4]{bi:C56}: ``{\em To a certain degree, however, the CM is sort of still in its infancy\,}.''

If connection proofs are expanded in the form of trees or DAGs (directed acyclic graph) as it is generally preferred in the tableaux-community, then the proof in Figure 1 expands to a tree in the usual way. In contrast to that the factorized proof expands to a DAG which turns out to be smaller than the corresponding tree. This already indicates the potential for smaller proofs and for more effective proof search behind this operation. Indeed, in the subsequent section we will see that factorization of formulas with multiplicities carries the potential for fundamental consequences in ATP.

Note that $\varphi_1 F_1^{\mu_1}=F_1^{\mu_1'}$. The proof for $F_1^{\mu_1}$ expands to a tree while that of $F_1^{\mu_1'}$ expands to a DAG, as just noted. That observation is the only reason why we have regarded the first one as the standard one. In view of the basic theorem underlying the CM (see eg.~\cite[6.4.C]{bi:B8}) this standard proof establishes the validity of the theorem principally in exactly the same way as does the proof for $F_1^{\mu_1'}$. Namely in both cases we select a multiplicity for the formula, then we further select for the formula with the selected multiplicity a spanning set of connections, and finally we show that there is a substitution which unifies each of these connections. Instead of focussing on the factorization relating the two formulas with multiplicities in our example, we could as well just speak of a less restricted application of the basic theorem underlying the CM in order to express the contribution of this note.

\section{A Conjecture along with Its Consequences}

It is well-known that the resolution method (RM) has an advantage over the tableaux method (TM) in that RM features cuts while TM does not. In~\cite[Sect.4]{bi:C56} we have introduced the notion $C_c$ to express this advantageous characteristic of RM. In consequence, there are theorems whose TM proofs are exponentially longer than their RM proofs with fatal consequences for searching for such proofs with TM. Most likely it is this advantage which so far has given RM the lead among its competitors.

Due to a version of the CM known as Elmar Eder's connection structure calculus (CSC)~\cite[Sect.~2.10]{be:A24} the CM shares this advantageous characteristic $C_c$ with RM. However, CSC proper so far has never been implemented. These facts lay the basis for the potential of the following conjecture.
\vspace{2mm}

{\bf Conjecture}. The connection method with factorization applied to formulas with multiplicities features the characteristic $C_c$. In addition, any connection proof with this factorization can linearly be transformed into a corresponding resolution proof, and vice versa.
\vspace{2mm}

The conjecture will not be proved in this short note, a task which is rather left to younger and active researchers in the field. But the long experience of the present author leads to his strong conviction of the conjecture's truth. Of course, any conjecture is not established without a precise proof. Everything what follows is to be understood under this proviso. Thus assume the conjecture could be proved correct.

First we point out that the following suggestions are worthwhile for consideration in any case, even if the assumption just made may fail to be correct. This is because the reduction by factorization of proof lengths and thus of the efforts of successful proof search is by itself advantageous without any doubt, as the example in the previous section illustrates.

As a little test for the conjecture, consider again the formula $F_1$ in Figure 1. Its shortest resolution proof consists of four resolution steps, each corresponding to one of the four connections in the connection proof with factorization, thus illustrating the conjecture in this simple case.

Factorization as described in the previous section may be integrated in a CM proof system in a straightforward manner. If this is carried out, the system will have achieved the same level of complexity as resolution. It has also achieved the same improvement that would be brought about by implementing the CSC. In fact, it seems rather obvious that the CSC may actually be interpeted in terms of factorization. Namely, the compacted form of matrices like that in~\cite[Fig.13,p.101]{be:A24} appears to be just an alternative way of dealing with multiplicities for the underlying formula with factorization. The expanded proof for this formula without considering factorization there is shown in~\cite[Fig.15,p.102]{be:A24}. Here the reduction is more impressive than the one for our example $F_1$, namely from multiplicity 8 down to 4.

While the CM with factorization may lead to much shorter proofs, as these examples illustrate, at the same time it inherits all the advantages which have made CM (and TM) attractive up to this day, despite the disadvantage of the present implementations in relation with $C_c$. These advantages have been described in numerous publications of the present and other authors. To mention here the simplest one: just compare the amount of information baggage required for the connection proof of $F_1$ (given formula, four connections with substitution) with that of its resolution proof (clause form of given formula, additionally four resolvents with involved substitutions).

In~\cite[Sect.3.1]{bi:C56} I have proposed to make sure ``{\em that any such common substructure occurs only once\,}'' in order to enhance CM systems. Factorization is the right tool to implement this very ambition. Namely, factorization as described above may lead to whole subproofs being factorized as a consequence of merging factorized literal occurrences. The example in Figure 1 is an extremely simple one for illustrative purposes and does therefore not demonstrate the whole potential of this general operation.

Of course, in the search for a proof it is not possible to know in advance the appropriate factorization $\varphi$. That is, factorization opens a further dimension in the search space for proofs. But taking this dimension into account will offer considerable benefits as well, as already discussed. So, one may, for instance, in a given proof task consider exploring various different $\varphi$'s, each combined with standard proof search. As an example for this option, recall $F_1$ in the previous section; there, the idea to factorize all possible connections of the form $(0,1^i)$, $i=1,2,\ldots$, beforehand is a rather obvious one. So, opting for this simplification at the outset, would result in the shorter proof without any further effort in this example. Considerations of this kind might lead to respective search strategies for more complicated cases.

Viewed in a different way and following the observation made in the last paragraph of the previous section, the search for a proof of a formula $F$ might be carried out -- indicated in rough terms -- by listing different multiplicities -- like $\mu_1$ and $\mu_1'$ in our example -- for the given formula $F$ to be proved and by determining for each of these the list of possible spanning sets of connections -- like the two sets in our example -- and test for each resulting connection structure whether or not there is a unifying substitution. The two listings involved in such an approach should of course not be done in a blind way, but with some reasonable, perhaps even learned strategy behind. All of this of course amounts to an extensive research for coming up with appropriate details.

Altogether, factorization as presented in this note opens a whole new domain of research in CM-oriented ATP. Given the improvement in terms of complexity as described in the conjecture, it is to be expected that CM systems enhanced by the new option will be able to challenge the performance of RM systems. When this would happen then it is very likely that it would lead to ATP changing its path of evolvement towards CM as its most successful proof method.

\bibliography{C:/Users/Bibel/Documents/Dateien/Beruf/bib/lit,C:/Users/Bibel/Documents/Dateien/Beruf/bib/my-publ}
\bibliographystyle{abbrv}

\end{document}